\documentstyle[twoside,12pt,epsf]{article}
\input epsf
\epsfverbosetrue

\def\singlespace {\smallskipamount=3pt plus1pt minus1pt
                  \medskipamount=6pt plus2pt minus2pt
                  \bigskipamount=12pt plus4pt minus4pt
                  \normalbaselineskip=12pt plus0pt minus0pt
                  \normallineskip=1pt
                  \normallineskiplimit=0pt
                  \jot=3pt
                  {\def\smallskip {\vskip\smallskipamount}}
                  {\def\medskip   {\vskip\medskipamount}}
                  {\def\bigskip   {\vskip\bigskipamount}}
                  {\setbox\strutbox=\hbox{\vrule
                    height8.5pt depth3.5pt width 0pt}}
                  \parskip 0pt
                  \normalbaselines}
\def\doublespace {\smallskipamount=6pt plus2pt minus2pt
                  \medskipamount=12pt plus4pt minus4pt
                  \bigskipamount=24pt plus8pt minus8pt
                  \normalbaselineskip=24pt plus0pt minus0pt
                  \normallineskip=2pt
                  \normallineskiplimit=0pt
                  \jot=6pt
                  {\def\smallskip {\vskip\smallskipamount}}
                  {\def\medskip   {\vskip\medskipamount}}
                  {\def\bigskip   {\vskip\bigskipamount}}
                  {\setbox\strutbox=\hbox{\vrule
                    height17.0pt depth7.0pt width 0pt}}
                  \parskip 12.0pt
                  \normalbaselines}
\def\halfspace {\smallskipamount=6pt plus2pt minus2pt
                  \medskipamount=12pt plus4pt minus4pt
                  \bigskipamount=24pt plus8pt minus8pt
                  \normalbaselineskip=16pt plus0pt minus0pt
                  \normallineskip=2pt
                  \normallineskiplimit=0pt
                  \jot=6pt
                  {\def\smallskip {\vskip\smallskipamount}}
                  {\def\medskip   {\vskip\medskipamount}}
                  {\def\bigskip   {\vskip\bigskipamount}}
                  {\setbox\strutbox=\hbox{\vrule
                    height17.0pt depth7.0pt width 0pt}}
                  \parskip 12.0pt
                  \normalbaselines}

\def\pprintspace {\smallskipamount=4pt plus1pt minus1pt
                  \medskipamount=9pt plus2pt minus2pt
                  \bigskipamount=16pt plus4pt minus4pt
                  \normalbaselineskip=14pt plus0pt minus0pt
                  \normallineskip=1pt
                  \normallineskiplimit=0pt
                  \jot=4pt
                  {\def\smallskip {\vskip\smallskipamount}}
                  {\def\medskip   {\vskip\medskipamount}}
                  {\def\bigskip   {\vskip\bigskipamount}}
                  {\setbox\strutbox=\hbox{\vrule
                   height9.5pt depth4.5pt width 0pt}}
                  \parskip 0pt
                  \normalbaselines}
\def\reidelspace {\smallskipamount=.1667 true in plus4pt minus2pt
                  \medskipamount=.3333 true in plus8pt minus2pt
                  \bigskipamount=13 true pt plus2pt minus2pt
                  \normalbaselineskip=13 true pt plus0pt minus0pt
                  \normallineskip=1 true pt
                  \normallineskiplimit=0 true pt
                  \jot=3pt
                  {\def\smallskip {\vskip\smallskipamount}}
                  {\def\medskip   {\vskip\medskipamount}}
                  {\def\bigskip   {\vskip\bigskipamount}}
                  {\setbox\strutbox=\hbox{\vrule
                    height8.5pt depth3.5pt width 0pt}}
                  \parskip 0pt
                  \normalbaselines}

\def\folio{\ifnum\pageno=1\nopagenumbers\else\number\pageno\fi}

\def\refitem{\par\noindent\hangindent 20pt}

\def\wisk#1{\ifmmode{#1}\else{$#1$}\fi}

\def\gt     {\wisk{>}}

\def\deg    {\wisk{^\circ}}
\def\ddeg   {\wisk{{\rlap.}^\circ}}

\def\thin{\thinspace}
\def\muK{\wisk{{\rm \mu K}}}

\begin{document}
\pagestyle{plain}
\pprintspace

\large
\begin{center}
Tests for Non-Gaussian Statistics in the DMR Four-Year Sky Maps
\end{center}

\medskip
\normalsize
\pprintspace
\noindent
\begin{center}
A.~Kogut\footnotemark[1]$^{,2}$,
A.J. Banday$^{1,3}$,
C.L. Bennett$^4$,
K. G\'{o}rski$^{1,5}$,
G. Hinshaw$^1$,
G.F. Smoot$^6$,
and
E.L. Wright$^7$
\end{center}
\footnotetext[1]{
~Hughes STX Corporation, Laboratory for Astronomy and Solar Physics, 
Code 685, NASA/GSFC, Greenbelt MD 20771. \newline
\indent~$^2$ E-mail: kogut@stars.gsfc.nasa.gov. \newline
\indent~$^3$ Current address: Max Planck Institut f\"{u}r Astrophysik,
85740 Garching Bei M\"{u}nchen, Germany. \newline
\indent~$^4$ Laboratory for Astronomy and Solar Physics, 
NASA Goddard Space Flight Center, Code 685, Greenbelt MD 20771. \newline
\indent~$^5$ On leave from Warsaw University Observatory,
Aleje Ujazdowskie 4, 00-478 Warszawa, Poland. \newline
\indent~$^6$ LBL, SSL, \& CfPA, Bldg 50-25,
University of California, Berkeley, CA 94720. \newline
\indent~$^7$ UCLA Astronomy, PO Box 951562, Los Angeles, CA 90095-1562. \newline
}

\medskip
\normalsize
\pprintspace
\begin{center}
{\it COBE} Preprint 96-07 \\
Submitted to {\it The Astrophysical Journal Letters} \\
January 5, 1996 \\
\end{center}


\medskip
\begin{center}
\large
ABSTRACT
\end{center}

\normalsize
\noindent
We search the high-latitude portion of the 
{\it COBE\thin}\footnotemark[8]
\noindent \footnotetext[8]{
~The National Aeronautics and Space Administration/Goddard Space Flight Center 
(NASA/GSFC) is responsible for the design, development, and operation of the
Cosmic Background Explorer ({\it COBE}).  
Scientific guidance is provided by the {\it COBE} Science Working Group.  
GSFC is also responsible for the analysis software 
and for the production of the mission data sets.
}
Differential Microwave Radiometers (DMR) 4-year sky maps 
for evidence of a non-Gaussian temperature distribution
in the cosmic microwave background.
The genus, 3-point correlation function,
and 2-point correlation function of temperature maxima and minima
are all in excellent agreement 
with the hypothesis that the CMB anisotropy
on angular scales of 7\deg ~or larger
represents a random-phase Gaussian field.
A likelihood comparison of the DMR sky maps to
a set of random-phase non-Gaussian toy models
selects the exact Gaussian model as most likely.
Monte Carlo simulations show that the 2-point correlation 
of the peaks and valleys in the maps 
provides the greatest discrimination among the class of models tested.

\noindent
{\it Subject headings:} 
cosmic microwave background --
cosmology: observations --
methods: statistical

\clearpage
\section{Introduction}
The angular distribution of the cosmic microwave background (CMB)
reflects the density distribution at the epoch of last scattering.
The statistical distribution of CMB temperature fluctuations
is thus an important probe of the initial conditions
for structure formation.  
Causally-connected regions at the surface of last scattering,
as viewed from the present epoch, subtend an angle $\sim 2\deg$.
Most of the current models for structure formation
predict the CMB temperature fluctuations
to be nearly Gaussian on larger angular scales.
In inflationary cosmologies, 
this Gaussian distribution reflects the
quantum origin of the density perturbations,
while in topological defect models,
it results from the superposition of a large number
of non-Gaussian perturbations at smaller angular scales
(the central limit theorem).

The CMB temperature field may be conveniently represented
in terms of spherical harmonics,
\begin{equation}
T(\theta,\phi) = \sum_{\ell m} a_{\ell m} Y_{\ell m}(\theta, \phi)
\label{harmonic_eq}
\end{equation}
and is said to be Gaussian
if the distribution each coefficient $a_{\ell m}$
follows a Gaussian probability distribution,
$$
P(a_{\ell m}) ~= ~\frac{1}{\sqrt{2 \pi \sigma_\ell^2}}
~\exp(- \frac{a_{\ell m}^2}{2 \sigma_\ell^2})
$$
with phases uniformly distributed in the range $[0, 2\pi]$.
Gaussian fields have the desirable property that they are 
completely specified by the power spectrum 
or its Fourier transform, the 2-point correlation function.
Thus, if the CMB is demonstrated to be a Gaussian field,
not only have current models passed an important consistency test,
but we are also spared the task of tabulating 
the full set of higher-order correlation functions.
Deviations from Gaussian statistics 
would necessitate a dramatic reappraisal of current models.

The {\it COBE} DMR experiment mapped the CMB at 7\deg ~angular resolution.
A number of authors have tested the DMR one- and two-year sky maps
and found good agreement with a Gaussian distribution
(Smoot et al.\ 1994;
Luo 1994;
Hinshaw et al.\ 1994, 1995;
Torres et al.\ 1995).
Kogut et al.\ 1995 compared the two-year DMR maps to non-Gaussian toy models
and found that the exact Gaussian model was the most likely descriptor
of the data.
In this {\it Letter} 
we compare the final 4-year DMR maps
(Bennett et al.\ 1996, Kogut et al.\ 1996)
to both Gaussian and non-Gaussian toy models
and discuss the relative ability of various tests to 
discriminate between the Gaussian and non-Gaussian models tested.

\section{Comparison with Gaussian models}
Cosmic variance, sample variance, and instrument noise
can combine to create non-Gaussian realizations
from initially Gaussian parent populations,
while the central limit theorem
pushes the CMB distribution back to Gaussian on large angular scales
(Scherrer \& Schaefer 1995).
We infer the statistical properties of the CMB parent population
using Monte Carlo techniques,
in which a certain statistic 
(3-point correlation function, extrema correlation function, or genus)
is calculated for many simulations of noisy CMB skies.
The simulations include all relevant aspects of the CMB model,
instrument noise, and data handling, 
including removal of the Galactic plane
and fitted dipole/quadrupole subtraction.
The same statistic is then calculated for the DMR data and
compared to the range observed in the simulations.
If the DMR data fall comfortably within the range of
equivalent Monte Carlo realizations, 
we accept the model as in agreement with the data; 
otherwise, we reject the model at some specified confidence level.

Unless otherwise noted,
we model the CMB as a scale-invariant Gaussian field 
$(P(k) ~\propto ~Q_{rms-PS}^2 ~k^n$ with $n=1$)
with quadrupole power-spectrum normalization
$Q_{rms-PS} = 18 ~\muK$
(G\'{o}rski et al.\ 1996),
evaluated at multipoles $2 \leq \ell \leq 40$
filtered through the DMR window function 
(Wright et al.\ 1994).
We compare these simulations to the
DMR 4-year 53 GHz (A+B)/2 sum map.
All statistics are evaluated only for the high-latitude sky
(3881 pixels at $|b| \gt 20\deg$
and custom cutouts near Ophiuchus and Orion,
Bennett et al.\ 1996)
from which a fitted dipole and quadrupole have been removed.

\subsection{3-Point Correlation Function}
A Gaussian field is fully characterized by its two-point correlation function.
All odd moments are identically zero, while all higher even moments
may be specified in terms of the 2-point function.
The 3-point correlation function and
its zero-lag value, the skewness,
are a first test for non-Gaussian signatures in the DMR data.

The general 3-point correlation function is the average product of 
three temperatures with a fixed relative orientation on the sky:
$$
C_3(\theta_1, \theta_2, \theta_3) ~= 
\langle T(\hat{n}_1)T(\hat{n}_2)T(\hat{n}_3)\rangle,
$$
where
$\hat{n}_1\cdot\hat{n}_2 = \cos\theta_1$,
$\hat{n}_2\cdot\hat{n}_3 = \cos\theta_2$, and
$\hat{n}_3\cdot\hat{n}_1 = \cos\theta_3$.
For computational reasons we evaluate only two special configurations,
the most sensitive of which is the pseudo-collapsed case
in which $\hat{n}_1$ is nearly parallel to $\hat{n}_2$.
Although the exact collapsed case $(\hat{n}_1 = \hat{n}_2)$
is the simplest configuration, 
it involves terms quadratic in the observed temperatures 
and thus suffers from a noise bias
(Hinshaw et al.\ 1994).
We avoid this problem by using the map geometry to our advantage.
Since the sky map pixels are 2\ddeg6 across 
and the DMR beam has a 7\deg ~FWHM, 
the nearest neighbor pixels will have correlated sky signal but
uncorrelated instrument noise.
We thus define
the pseudo-collapsed 3-point correlation function as
$$
C_3^{\rm (pc)} = \sum_{i,j,k} w_i w_j w_k\ T_i T_j T_k / \sum_{i,j,k} w_iw_jw_k
$$
where $T_i$ is the temperature in pixel $i$,
$w$ is the pixel weight,
the sum on $j$ runs over all pixels that are nearest neighbors to $i$,
and the sum on $k$ is over all pixels (except $j$) 
within a fixed angular separation of $i$.
We also evaluate the equilateral configuration,
$C_3^{(e)}$, for which $ \theta_1 = \theta_2 = \theta_3$.
Since most pixels have 8 nearest neighbors, 
the pseudo-collapsed configuration
sums over 8 times as many pixel triples as the equilateral configuration,
resulting in a significant increase in sensitivity.

\begin{figure}[t]
\epsfxsize=6.0truein
\epsfbox{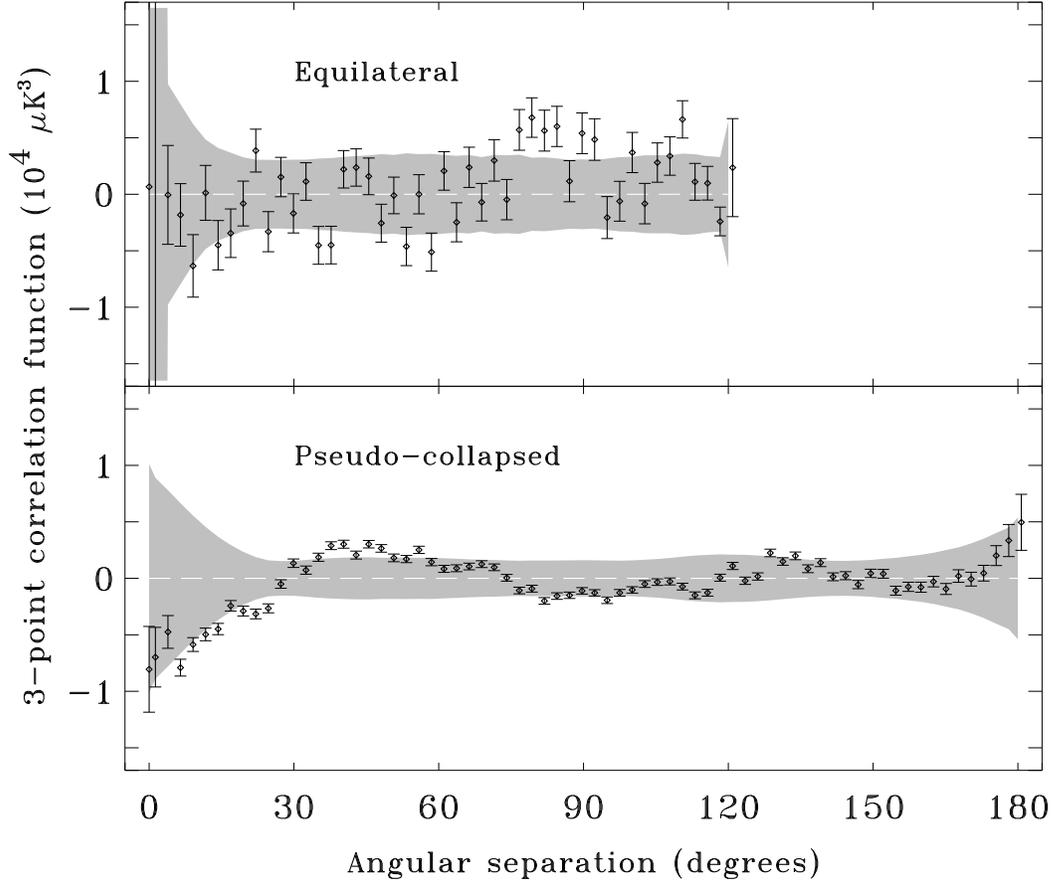}
\caption{
Three-point correlation function
of the 4-year 53 GHz (A+B)/2 sum map,
compared to simulated maps with 
$Q_{rms-PS} = 18 ~\muK$, $n=1$, and 53 GHz noise properties
(thermodynamic temperature units).
The error bars represent the uncertainty due to instrument noise alone,
while the gray bands represent the 68\% confidence region
from the combined noise and Gaussian sky signal.
(top) Equilateral triangle configuration.
(bottom) Pseudo-collapsed configuration.
The fluctuations about zero correlation
are too large to be ascribed to instrument noise alone,
but are consistent with the range of fluctuations
expected from a Gaussian CMB.
}
\label{3pt_fig}
\end{figure}

Figure \ref{3pt_fig} shows the pseudo-collapsed and equilateral
3-point correlation functions for the 4-year DMR 53 GHz (A+B)/2 maps,
evaluated with unit pixel weight ($w=1$) in 2\ddeg6 angular bins.
The gray band in each plot shows the 68\% confidence region
derived from 2000 random realizations of scale-invariant CMB anisotropy
superposed with instrument noise, 
and includes the effects of cosmic variance.
Deviations from zero in $C_3$ for either configuration
are well within the range expected for a single realization 
of a stochastic process.
We quantify this using a $\chi^2$ statistic,
\begin{equation}
\chi^2 ~= ~\sum_{\alpha \beta} 
~(D_\alpha - \langle S_\alpha \rangle)
~({\bf M}^{-1})_{\alpha \beta} 
~(D_\beta - \langle S_\beta \rangle)
\label{chisq_eq}
\end{equation}
where $D_\alpha$ is the DMR 3-point correlation function,
$\langle S_\alpha \rangle$ is the mean correlation function from the 
simulations (in this case identically zero),
and
\begin{equation}
{\bf M}_{\alpha \beta} ~= ~\frac{1}{N} ~\sum_i 
~(S^{(i)}_\alpha - \langle S_\alpha \rangle)
~(S^{(i)}_\beta - \langle S_\beta \rangle)
\label{covar_eq}
\end{equation}
is the mean covariance matrix 
between the angular bins $\alpha$ and $\beta$
computed from the $N=2000$ correlation functions $S^{(i)}$ in the simulations.
We calculate 2000 such $\chi^2$ values 
by replacing the 3-point correlation function from the DMR data, $D_\alpha$,
with the correlation function from the $i$th simulation, $S^{(i)}_\alpha$ .
The DMR maps fall near the median of the resulting distribution:
66\% of the simulations had larger $\chi^2$ 
than the DMR pseudo-collapsed 3-point correlation function,
while 31\% of the simulations had larger $\chi^2$ 
than the DMR equilateral configuration.
The 3-point correlation function shows no evidence for
a non-Gaussian CMB distribution.

\subsection{Genus}
Topology provides another test of the Gaussian hypothesis.
The statistical properties of the CMB can be characterized
by the excursion regions enclosed by isotemperature contours.
The genus is the total curvature of the contours 
at fixed temperature, and may loosely be defined as the
number of isolated high temperature regions (hot spots) minus the
number of isolated low temperature regions (cold spots).
For a random Gaussian field, the genus per unit area is 
$$
g(\nu) ~= ~\frac{1}{(2 \pi)^{3/2} ~\theta_c^2} ~\nu ~e^{-\nu^2/2}
$$
(Gott et al.\ 1990),
where $\nu$ is the temperature contour threshold in units of the
standard deviation $\sigma$ of the field, and
$\theta_c$ is the correlation angle, related to the 2-point correlation 
function $C_2(\theta)$ by
$$
\theta_c^{-2} = -{1 \over C_2(0)} {d^2 C_2 \over d \theta^2}|_{\theta=0}.
$$
The genus per unit area is a locally invariant quantity
and is insensitive to incomplete sky coverage 
(e.g. removal of the Galactic plane).

The genus of a Gaussian field depends on two parameters:
the correlation angle $\theta_c$ and the standard deviation $\sigma$.
The DMR noise is nearly uncorrelated (Lineweaver et al.\ 1994);
its correlation angle is determined by the pixelization.
The correlation angle of a scale-free CMB depends only on the index $n$.
For a noisy map, however,
both the correlation angle and the standard deviation
depend on the relative amplitude of the signal and noise terms
and hence acquire a dependence on the CMB amplitude normalization $Q_{rms-PS}$.
Since $\theta_{c,{\rm noise}} \ll \theta_{c,{\rm CMB}}$,
the sensitivity of the genus to the CMB terms can be enhanced
by smoothing the map,
provided the smoothing dispersion remains less than $\theta_{c,{\rm CMB}}$.
We evaluate the genus at 3 smoothing angles
with full width at half maximum 
0\deg ~(no smoothing), 5\deg, and 10\deg.
At each smoothing angle, a nearest-neighbor algorithm
counts the number of hot and cold spots 
(contiguous pixels above or below the temperature threshold)
at 11 thresholds $\nu$ 
from -$2.5\sigma$ to +$2.5\sigma$ in steps of $0.5\sigma$.
We adopt the difference in spot number as an estimate of the genus
and use the Monte Carlo simulations to calibrate any difference 
induced by the Galactic cut
between this definition and the total curvature.
The two definitions are identical in the absence of a Galactic cut,
and the nearest-neighbor algorithm allows significant computational savings.

\begin{figure}[t]
\epsfxsize=6.0truein
\epsfbox{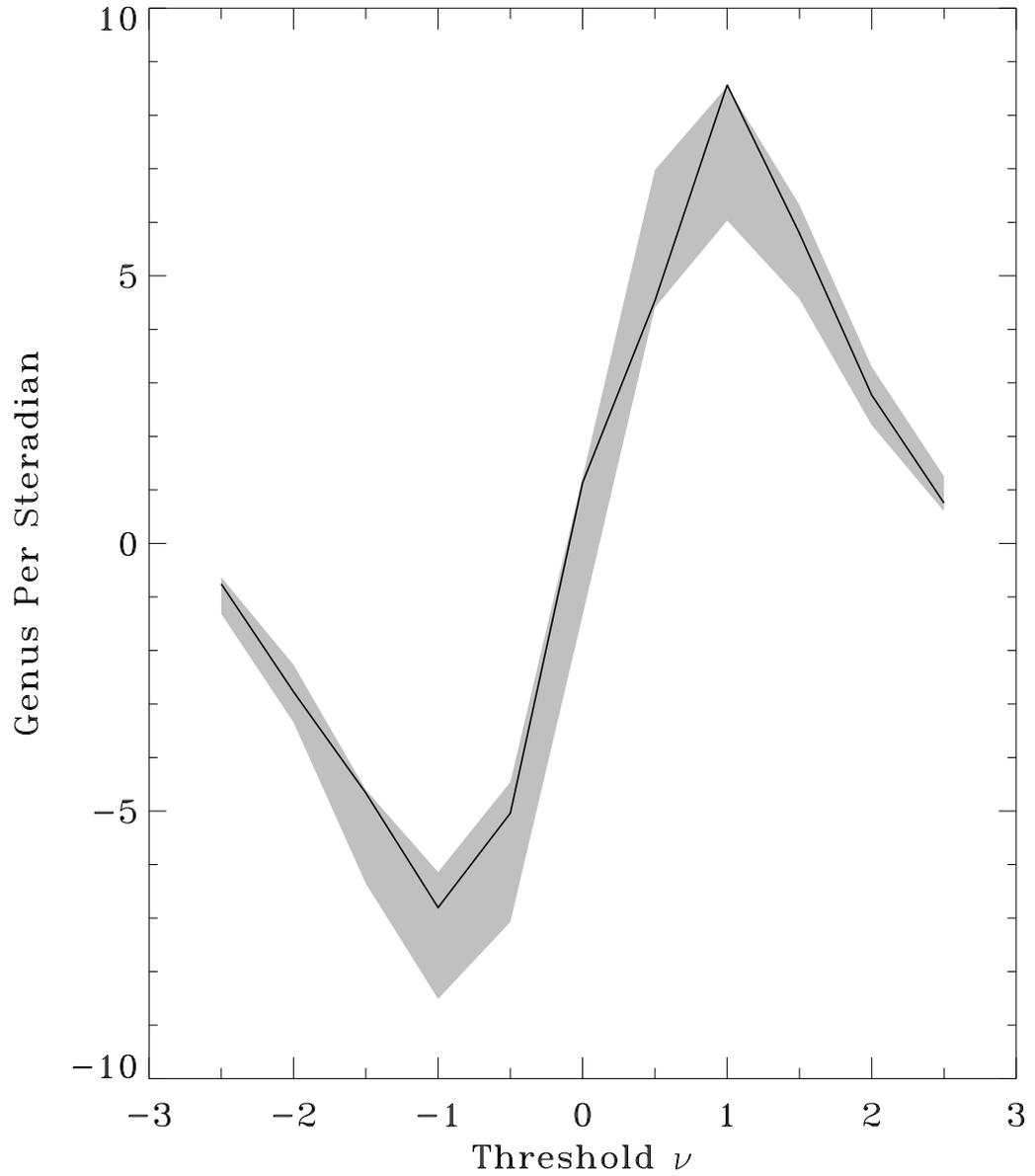}
\caption{
Genus curve for the 4-year 53 GHz (A+B)/2 sum map,
compared to simulated maps with 
$Q_{rms-PS} = 18 ~\muK$, $n=1$, and 53 GHz noise properties.
The gray band represents the 68\% confidence region from the simulations.
The genus is in excellent agreement with the Gaussian CMB model.
}
\label{genus_fig}
\end{figure}

Figure \ref{genus_fig} shows the genus of the 53 GHz (A+B)/2 map
smoothed with a Gaussian of 5\deg ~full width at half maximum (FWHM).
The DMR genus is well within the range found by the simulations.
As with the 3-point correlation function, we quantify this 
using a $\chi^2$ statistic
(equations \ref{chisq_eq} and \ref{covar_eq}
with vectors $D_\alpha$ and $S_\alpha$ 
now consisting of the genus 
in $k=33$ bins for the 11 thresholds and 3 smoothing angles 
evaluated simultaneously).
The DMR genus is near the median of the Gaussian models:
51\% of the simulations had larger $\chi^2$.

Since the instrument noise properties are known,
we may interpret the genus in terms of the CMB model parameters.
We evaluate the $\chi^2$ of the DMR genus on a
2-dimensional grid in $Q_{rms-PS}$ and $n$,
use the value at each grid point
to determine the likelihood
$$
{\cal L}(Q_{rms-PS},n) ~= ~(2 \pi)^{-k/2}
~{ {\rm exp}(- \frac{1}{2} \chi^2) \over \sqrt{ {\rm det}({\bf M}) } },
$$
and search for the maximum in the resulting likelihood distribution.
We also perform a probabilistic search by converting the $\chi^2$
at each grid point to a probability $P(\chi^2)$,
defined as the fraction of simulations with $\chi^2$
larger than the DMR value at that grid point.
We calibrate both methods and search for biases 
by generating 2000 simulations with fixed parameters $Q_{rms-PS}$ and $n$,
fitting each realization,
and examining the ``cloud'' of fitted values.
The $\chi^2$ probability test has a small bias of 0.8 \muK
~toward smaller values of $Q_{rms-PS}$;
the maximum likelihood test is unbiased.

Both methods yield consistent results on the DMR 53 GHz (A+B)/2 data.
The likelihood shows a strong ridge
along the line
$Q_{rms-PS} ~= ~(17.3 \pm 1.8) ~- ~(8.8 \pm 0.2)(n-1)$ \muK
~with likelihood maximum at
$Q_{rms-PS} = 24.5 \pm 5.0 ~\muK$ and $n = 0.2 \pm 0.6$
(68\% confidence interval).
The value at the maximum has $P(\chi^2)$=77\%,
and is not statistically preferred over the more precise value
derived from the power spectrum of the combined DMR maps
(G\'{o}rski et al.\ 1996).

\subsection{Extrema Correlation Function}
Another test of the statistical distribution
is the angular correlation of the extrema (peaks and valleys)
in the temperature field,
defined as those points for which $\nabla T = 0$.
For a pixelized map, this reduces to the collection of pixels
hotter or colder than all of their nearest neighbors.
Specifying pixels hotter than their neighbors
produces a set of ``peaks'',
while specifying colder pixels produces ``valleys''.

The number density of peaks or valleys is dominated
by the noise properties of the map (Kogut et al.\ 1994).
The clustering of the extrema,
as measured by the 2-point correlation function of the maxima and minima,
provides additional information on the underlying CMB temperature field.
We define the extrema correlation function as
$$
C_{\rm ext}(\theta) ~=
~{ {\sum_{i,j} w_i w_j T_i T_j} \over {\sum_{i,j} w_i w_j} }
$$
(Kogut et al.\ 1995),
where the sum over pixel temperatures is restricted to the
set of maxima and minima pixels.
Bond \& Efstathiou (1987) provide analytic approximations for
$C_{\rm ext}(\theta)$ for random Gaussian fields
but do not explicitly include the effects of instrument noise.
Since the correlation properties of the non-uniform noise in the DMR maps
are different from the underlying CMB temperature field,
we smooth the maps with a 7\deg ~FWHM Gaussian
prior to collating the extrema
as a compromise between
suppressing noise and removing CMB power at small scales.
We use Monte Carlo techniques as described above
to calibrate the resulting distribution of simulated correlation functions.

The clustering of extrema depends on the threshold $\nu$.
We evaluate $C_{\rm ext}(\theta)$ at thresholds $|\nu|$ = 0, 1, and 2.
Since, by definition, two peaks can not be adjacent,
we ignore both the bin at zero angular separation and the first non-zero bin
in all subsequent analysis.
Simulations show that the results are dominated by the 
first few remaining bins; consequently,
we speed processing by truncating the correlation function
at separation $\theta = 60\deg$ for a total of 22 angular bins
at each of the three thresholds.

The angular distribution of peaks and valleys in the DMR maps
is in excellent agreement with Gaussian statistics.
We calculate a $\chi^2$ value for the DMR extrema correlation function
and compare it to 2000 simulations of CMB models with
$Q_{rms-PS} = 18 ~\muK$ and $n=1$.
70\% of the simulations had larger $\chi^2$ than the DMR maps.

\section{Comparison with non-Gaussian models}
\begin{figure}[t]
\epsfxsize=6.0truein
\epsfbox{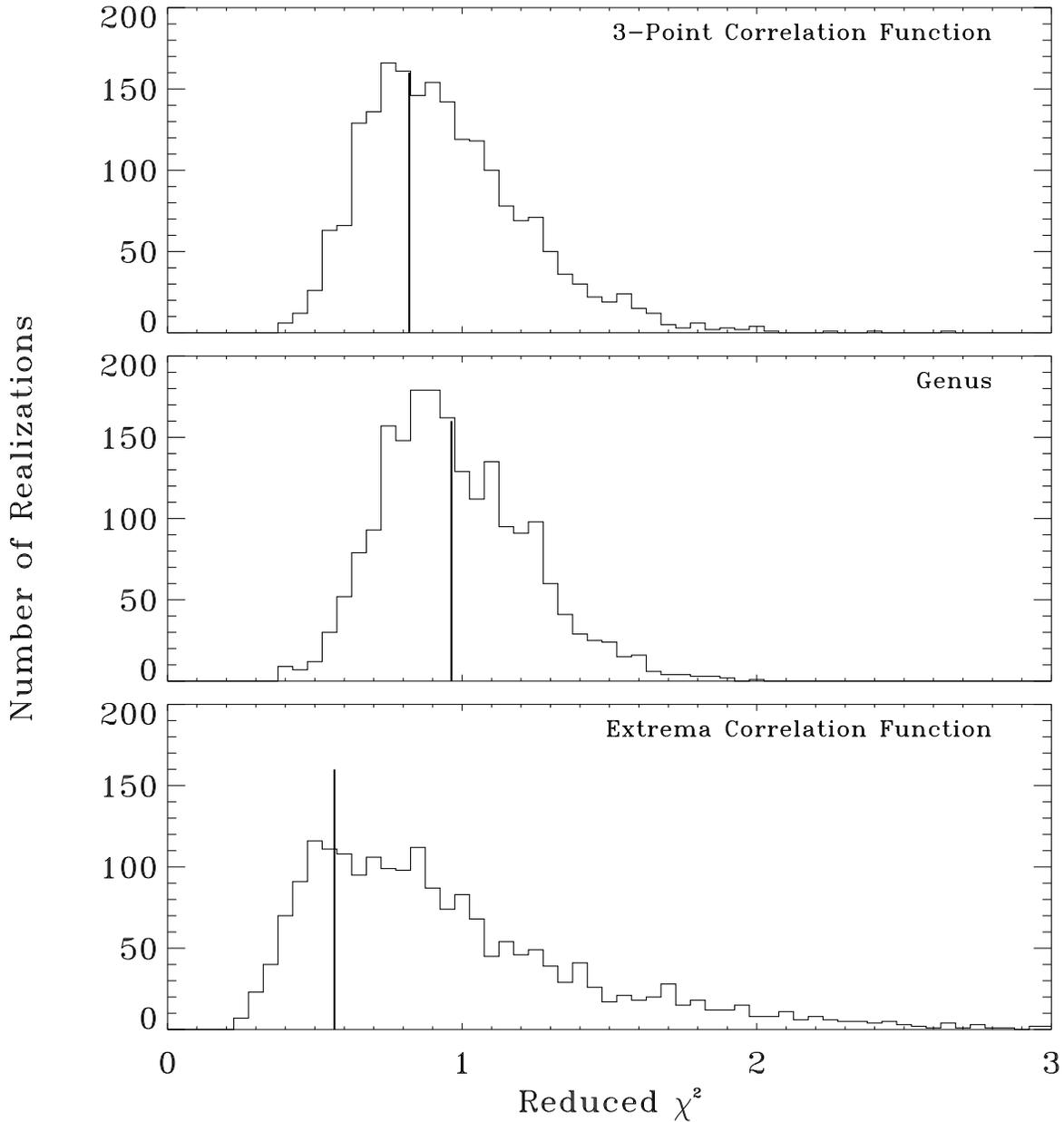}
\caption{
$\chi^2$ of the DMR maps compared to the distribution
derived from simulations with the Gaussian CMB model,
divided by the number of bins used to evaluated the $\chi^2$.
(top) Three-point correlation function.
(middle) Genus.
(bottom) Extrema correlation function.
The $\chi^2$ values from the DMR data (vertical bars)
are near the median of the distribution 
expected from Gaussian CMB anisotropy.
}
\label{chisq_fig}
\end{figure}

Figure \ref{chisq_fig} summarizes the comparison of the 
53 GHz (A+B)/2 four-year sky maps 
with Gaussian CMB models.
The $\chi^2$ derived from the
3-point correlation function, genus, and extrema correlation function
of the DMR data 
fall near the median of the $\chi^2$ distributions
derived from the simulations:
the DMR data are in excellent agreement with the hypothesis that
the CMB is a realization of a parent population
of Gaussian density perturbations.

Given the large DMR beam, the agreement of the DMR data with Gaussian models
does not necessarily place tight limits
on {\it non}-Gaussian models.
There are an infinite number of non-Gaussian distributions,
from which we wish to select those representative 
of physically interesting physical processes.
Ideally, we would use Monte Carlo realizations
of specific cosmological models such as texture or global monopoles.
Unfortunately, no analytic expressions exist for the temperature anisotropy
produced by these models; generating a single realization
requires a substantial computational investment
to follow the causal field ordering in some large volume.
Instead, we will modify the standard Gaussian model
to test generic but interesting non-Gaussian amplitude distributions.

We use a likelihood analysis of the statistical tests described above
to compare the DMR maps against a set of toy models
for which the spherical harmonic amplitudes $a_{\ell m}$ 
(Eq. \ref{harmonic_eq})
are random variables with zero mean
drawn from $\chi^2_N$ parent populations
with $N$ = 1, 5, 15, 30, or 60 degrees of freedom.
We fix the variance $\sigma_\ell^2$ of the parent population
at the same value as the standard Gaussian model,
$$
\sigma_\ell^2 ~= ~(Q_{rms-PS})^2 ~{4 \pi \over 5}
~{\Gamma[l+(n-1)/2] ~\Gamma[(9-n)/2] \over \Gamma[l+(5-n)/2] \Gamma[(3+n)/2] }
$$
(Bond \& Efstathiou 1987),
providing a simple, computationally efficient modification
to the standard Gaussian model.
Although the non-Gaussian {\it amplitude} distributions tested here are
skew-positive,
the resulting {\it temperature} distributions (sky maps)
are the convolution of the $a_{\ell m}$
with the spherical harmonics $Y_{\ell m}$
and are thus characterized not by the skewness
but by a positive kurtosis in the
distribution of temperatures $T$
(e.g., higher ``wings'' than a Gaussian distribution).
The toy models provide a smooth transition 
from strongly non-Gaussian ($N$=1) to nearly Gaussian ($N$=60),
and may be compared to specific models of structure formation
(Weinberg \& Cole 1992, Kogut et al.\ 1995).
Cosmological models with rare high-amplitude peaks,
typified by topological defect models such as texture or global monopoles
can be compared to the $\chi^2$ models with few degrees of freedom,
which tend to produce such features.

\begin{figure}[t]
\epsfxsize=6.0truein
\epsfbox{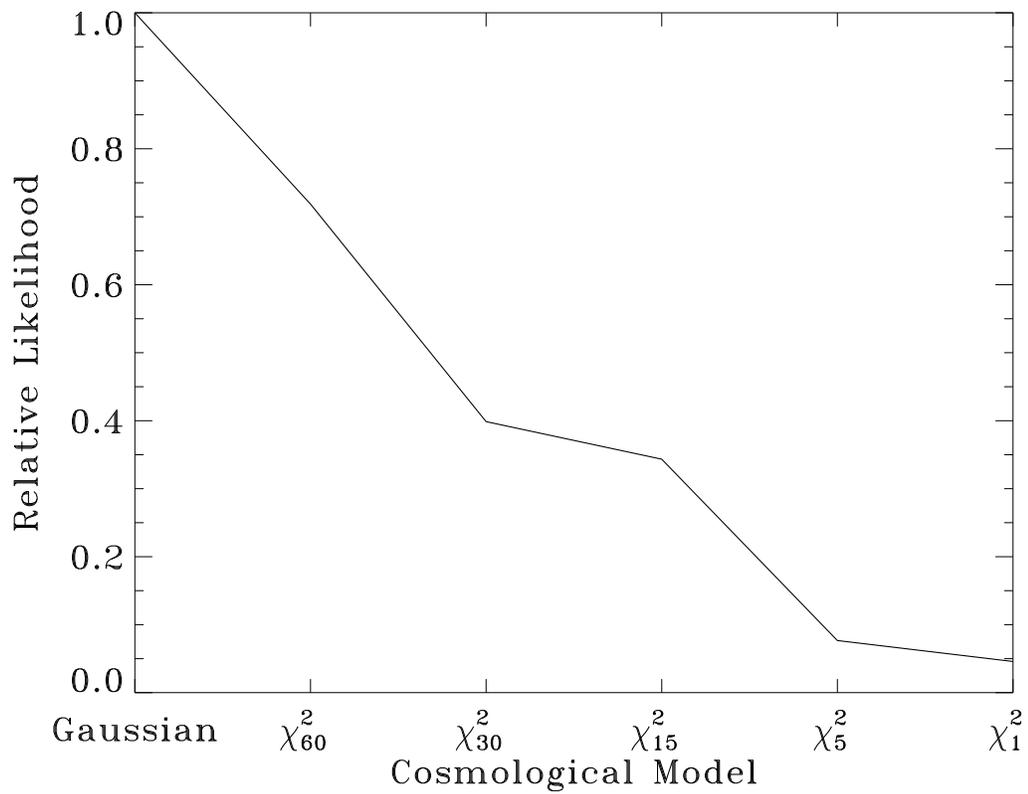}
\caption{
Likelihood function of the 4 year DMR 53 GHz (A+B)/2 
extrema correlation function 
(marginalized over the CMB normalization $Q_{rms-PS}$)
for Gaussian and non-Gaussian models.
}
\label{like_fig}
\end{figure}

Figure \ref{like_fig} shows the likelihood of the 4-year DMR maps
for the Gaussian and $\chi^2_N$ models,
derived from the extrema correlation function.
The likelihood is greatest for the exact Gaussian model,
and decreases as the toy models become increasingly non-Gaussian.
The Gaussian model remains the most likely descriptor of the data
when the genus replaces the extrema correlation in the likelihood analysis.

Given that the DMR data ``prefer'' the Gaussian model,
how confident are we that the CMB is not well described by
one of the $\chi^2$ toy models?
Since cosmic variance can produce a nearly Gaussian sky
from a non-Gaussian parent population and vice-versa,
marginalizing over model parameters
can produce misleading confidence intervals
(see the discussion of this point in Kogut et al.\ 1995).
We quantify the ability of each statistical test
to discriminate between Gaussian and non-Gaussian models 
using Monte Carlo simulations.
We take 2000 simulations from each of the models in turn,
and for each simulation
generate the likelihood function ${\cal L}(Q_{rms-PS},N)$.
We then count the number of times the Gaussian model is selected 
as ``most likely.''
We obtain the DMR result
(Gaussian model chosen as most likely)
five times more often when the input model is Gaussian 
than when the input is any of the non-Gaussian models.
The method is unbiased: if we ask how often the $\chi^2_{15}$ model 
would be selected as most likely, we find the greatest probability
if the input model is $\chi^2_{15}$, and so forth.
In this posterior sense,
the CMB is five times more likely to be a realization of a Gaussian process
than any of the $\chi^2$ models tested.
Table \ref{prob_table} gives the relative probability
to observe a Gaussian sky 
if the CMB is a sample drawn from each of the toy models,
derived using either the genus or the extrema correlation function
(since all of the models tested have zero skewness,
the 3-point correlation function does not discriminate among them).
The extrema correlation function discriminates among the toy models
twice as well as the genus.

\section{Conclusions}
We have tested the statistical distribution of temperature fluctuations
in the DMR 4-year 53 GHz (A+B)/2 summed sky map
against both Gaussian and non-Gaussian models of CMB anisotropy
using three statistical tests: 
the 3-point correlation function, genus, and 2-point correlation function
of temperature maxima and minima.
A goodness-of-fit test using the $\chi^2$ of each function
shows the DMR data to lie near the median
of the distribution of $\chi^2$ values
evaluated for randomly generated Gaussian models
with amplitude normalization $Q_{rms-PS} = 18 ~\muK$ and
power-law index $n=1$.
The DMR data are thus in excellent agreement with the hypothesis
that the observed CMB anisotropy
represents a realization of a Gaussian parent process.
We test an alternate hypothesis,
that the CMB represents a realization of a non-Gaussian process,
by evaluating the likelihood of the DMR data
against a grid of non-Gaussian toy models
whose spherical harmonic coefficients $a_{\ell m}$
are drawn from $\chi^2_N$ distributions 
with $N$ = 1, 5, 15, 30, or 60 degrees of freedom.
The likelihood reaches a maximum for the exact Gaussian model.
Cosmic variance produces significant cross-talk 
among the class of models tested
and precludes the simple identification of confidence intervals on $N$.
Simulations in which a similar likelihood analysis is performed 
on random realizations of either Gaussian or non-Gaussian skies
demonstrates that the method correctly identifies the input model
with a $\sim$15\% error rate against any of the other models;
the error rate is nearly independent of $N$.
The DMR result 
(Gaussian model selected as most likely)
occurs five times more often when the input to the simulations
is, in fact, Gaussian,
than when it is any of the non-Gaussian toy models.
We thus conclude that,
not only do Gaussian power-law models 
provide adequate description of the CMB anisotropy on large angular scales,
but that the non-Gaussian models tested
are five times less likely to describe the true statistical distribution.

\vspace{0.5 in}
We gratefully acknowledge the dedicated efforts
of the many people responsible for the {\it COBE} DMR data:
the NASA office of Space Sciences,
the {\it COBE} flight team,
and all of those who helped process and analyze the data.

\clearpage
\begin{center}
\large
{\bf References}
\end{center}

\refitem
Bennett, C.L., et al.\ 1996, ApJ Letters, submitted

\refitem
Bond, J.R., and Efstathiou, G., 1987, MNRAS, 226, 655

\refitem
Gott, J.R., Park, C., Juskiewcz, R., Biew, W.E., Bennett, D.P,
Bochet, F.R., \& Stebbins, A. 1990, ApJ, 352

\refitem
G\'{o}rski, K.M., Banday, A.J., Bennett, C.L., Hinshaw, G., Kogut, A.,
Smoot, G.F., \& Wright, E.L.\ 1996, ApJ Letters, submitted

\refitem
Hinshaw, G., et al.\ 1994, ApJ, 431, 1

\refitem
---, Banday, A.J., Bennett, C.L., G\'{o}rski, K.M., \& Kogut, A. 
1995, ApJ, 446, L67

\refitem
Kogut, A., Banday, A.J., Bennett, C.L., Hinshaw, G., Loewenstein, K.,
Lubin, P., Smoot, G.F., and Wright, E.L., 1994, ApJ, 433, 435

\refitem
---, Banday, A.J., Bennett, C.L., Hinshaw, G., Lubin, P.M., 
\& Smoot, G.F. 1995, ApJ, 439, L29

\refitem
---, et al.\ 1996, ApJ, submitted

\refitem
Lineweaver, C.H., et al.\ 1994, ApJ, 436, 452

\refitem
Luo, X. 1994, Phys. Rev. D., 49, 3810

\refitem
Scherrer, R.J., \& Schaefer, R.K 1995, ApJ, 446, 44

\refitem
Smoot, G.F., Tenorio, L., Banday, A.J., Kogut, A., Wright, E.L.,
Hinshaw, G., \& Bennett, C.L. 1994, ApJ, 437, 1

\refitem
Torres, S., Cay\'{o}n, L., Mart\'{i}nez-Gonz\'{a}lez, E.,
\& Sanz, J.L. 1995, MNRAS, 274, 853

\refitem
Weinberg, D.H., \& Cole, S. 1992, MNRAS, 259, 652

\refitem
Wright, E.L., et al.\ 1994, ApJ, 420, 1


\vfill
\clearpage

\normalsize
\halfspace
\begin{table}
\caption{\label{prob_table}
Relative Probability To Select Gaussian Model$^a$}
\begin{center}
\begin{tabular}{l c c}
\hline
Input & \multicolumn{2}{c}{Statistical Test} \\
Model & Extrema Correlation & Genus \\
\hline
Gaussian       & 5.1 & 2.7 \\
$\chi^2_{60}$  & 1.1 & 1.0 \\
$\chi^2_{30}$  & 1.0 & 1.2 \\
$\chi^2_{15}$  & 1.2 & 1.2 \\
$\chi^2_{5}$   & 1.1 & 1.3 \\
$\chi^2_{1}$   & 1.1 & 1.1 \\
\hline
\end{tabular}
\end{center}
$^a$~Number of realizations from each input model for which the Gaussian model
was selected as most likely, divided by the smallest such value to yield
the relative probability of obtaining the DMR result from each input model
(see text).
\end{table}

\clearpage

\clearpage

\clearpage

\clearpage

\end{document}